\documentclass[aps,prl,twocolumn,showpacs,longbibliography]{revtex4-2}
\usepackage{physics}
\usepackage{tikz} 
\usetikzlibrary{shapes,arrows,positioning,automata,backgrounds,calc,er,patterns}
\usepackage{graphicx}
\usepackage{dcolumn}
\usepackage{bm}
\usepackage{amsmath}
\usepackage{amssymb}
\usepackage{amsthm}
\usepackage{amsfonts}
\usepackage{enumerate}
\usepackage{siunitx}
\usepackage{latexsym}

\usepackage[colorlinks=true,urlcolor=blue,citecolor=blue,linkcolor=blue]{hyperref}

\definecolor{purple}{rgb}{0.8,0,0.6}
\definecolor{darkgreen}{rgb}{0.00,0.6,0.00}

\newcommand{\affiliationSICNU}{Department of Physics, Institute of Solid State Physics and Center for Computational Sciences, Sichuan Normal University, Chengdu, Sichuan 610066, China}

\begin{document}

	\title{Nonlinear opto-magnetic signature of $d$-wave altermagnets}
	\author{Lijun Yang}
\author{Long Liang}\email{longliang@sicnu.edu.cn}
\affiliation{\affiliationSICNU}

\begin{abstract}
	Altermagnetism, a recently discovered collinear magnetic order with net zero magnetization but exhibiting spin-splitting band structure, has attracted much research interest due to the rich fundamental physics and possible applications. 
	In this work, we investigate the opto-magnetic response of $d$-wave altermagnets, focusing on the inverse Cotton-Mouton effect\textemdash the induction of  static magnetization via  linearly polarized light. We find that 
	the direction of the induced magnetization is determined by the N\'eel vector. Moreover, its  magnitude exhibits a periodic dependence on the polarization angle of the incident light,  a hallmark of the system's symmetry.   
	 {Our findings demonstrate that the inverse Cotton-Mouton effect offers a direct method both for detecting $d$-wave altermagnets and for probing their intrinsic properties.}
\end{abstract}
\maketitle

\emph{Introduction.\textemdash}Altermagnetism~\cite{PhysRevX.12.011028,PhysRevX.12.031042,PhysRevX.12.040501,Bai_2024} has recently been proposed as  a new type of magnetic order  that goes beyond conventional ferromagnetic and antiferromagnetic orders.
Altermagnets  possess fully compensated collinear magnetic moments like conventional antiferromagnets. However, the sublattices of antiparallel spins are connected by nontrivial rotational symmetry rather than by translation or inversion symmetry. 
This results in a momentum-dependent spin-splitting of the electronic band structure~\cite{PhysRevB.75.115103,C5CP07806G,10.7566/JPSJ.88.123702,PhysRevB.102.014422,PhysRevMaterials.5.014409}, reminiscent of ferromagnets, even in the absence of spin-orbit coupling.
The spin-splitting changes sign across the Brillouin zone,
as dictated by  
the altermagnetic order parameter classified by the spin group~\cite{PhysRevX.12.021016,PhysRevX.14.031038,PhysRevX.14.031037,PhysRevX.14.031039,10.21468/SciPostPhys.18.3.109}. A Landau theory for altermagnetism has been developed~\cite{PhysRevLett.132.176702}.

The spin-split band structure has been observed using spin and angle 
resolved photoemission spectroscopy  in several altermagnetic material candidates~\cite{PhysRevLett.132.036702,Krempasky2024,Zhu2024,PhysRevLett.133.206401,Jiang2025,Zhang2025}.
The large spin-momentum interaction in altermagnets enables various phenonema, such as spin current generation~\cite{Naka2019,PhysRevLett.126.127701,Bose2022}, anomalous Hall  effect~\cite{10.1126/sciadv.aaz8809,10.1073/pnas.2108924118,Feng2022,PhysRevB.110.094425}, 
 piezomagnetic effect~\cite{Ma2021,PhysRevMaterials.8.L041402,ZhuYu2024,PhysRevX.14.011019}, and chiral magnons~\cite{PhysRevLett.131.256703,PhysRevLett.133.156702,cichutek2025,fgc1-5blp}.

The optical control of magnetization is a major goal of next-generation spintronics and  information technology~\cite{RevModPhys.82.2731,Saidl2017,RevModPhys.90.015005,Nemec2018}.
While conventional antiferromagnets offer tantalizing advantages, such as zero stray fields and ultrafast dynamics, their manipulation with light remains a significant challenge
~\cite{RevModPhys.90.015005,Nemec2018}. 
Altermagnets provide novel possibilities for optical control by  combining the advantages of ferromagnets and antiferromagnets. 
The optical manipulation of altermagnets, however, remains a largely unexplored frontier~\cite{weber2024optical,KIMEL2024172039,Adamantopoulos2024,weber2024ultrafast,zhou2025magnetizingaltermagnetsultrafastasymmetric,PhysRevB.111.L201405,PhysRevB.111.064428,bzzy-ngcs,yjp4-gkj9}.

\begin{figure}
	\includegraphics[angle=0,width=\linewidth]{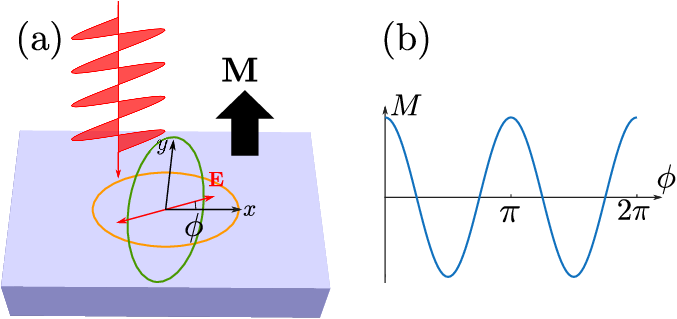}
	\caption{(a) Schematic setup of the inverse Cotton-Mouton effect in a $d$-wave altermagnet. 
		Linearly polarized light incident on the sample induces a static magnetization. {The light is incident from the $z$-direction and polarized in the $x-y$ plane.} The two ellipses represent the spin-split Fermi surfaces for spin-up and spin-down electrons. The magnitude of the induced magnetization depends on the polarization angle $\phi$ of the light, while its direction is governed by the N\'eel vector. As demonstrated in panel (b), the resulting magnetization exhibits a $\pi$-periodic angular dependence. The induced magnetization can be detected by magneto-optical Kerr rotation  or SQUID.
	}
	\label{Fig:fig1}
\end{figure}

In this work, we study the nonlinear opto-magnetic response of $d$-wave altermagnets.  
We focus on the  inverse Cotton-Mouton effect (ICME)~\cite{Pershan-Malmstrom:1966,1984Shen,PhysRevLett.99.167205,PhysRevLett.123.157202}, which describes the static magnetization induced by linearly polarized light. The corresponding experimental setup is schematically shown in Fig.~\ref{Fig:fig1}. Symmetry analysis shows that the magnetization depends on the direction of polarization, which reflects the symmetry of altermagnets. 
Using the 
Keldysh formalism, we present a microscopic theory of  the  effect. We find that  
the direction of  magnetization is along the direction of the N\'eel order, thus providing an optical method to detect the N\'eel order in altermagnets.
We estimate the order of magnitude of the induced magnetization for altermagnet candidate KRu$_4$O$_8$~\cite{PhysRevX.12.031042} and show its frequency and temperature dependence. Our results demonstrate the ICME as a mechanism for detecting  the unique properties of altermagnet.

\emph{Symmetry considerations.\textemdash}
The static magnetization induced by  monochromatic light with frequency $\Omega$ can be written as 
\begin{align}
	M_i=\chi_{ijl}(\Omega)E_j(\Omega)E_l(-\Omega),
\end{align}
where the Einstein summation convention is used. 
The response function $\chi_{ijl}$ is a third order pseudotensor. In contrast to the linear response, this second-order effect  is allowed in systems with inversion symmetry. 
When the first index is fixed, the response function $\chi_{ijl}$ can be represented as a matrix. Its antisymmetric part describes the inverse Faraday effect~\cite{Pitaevskii:1960,Pershan:1963,Pershan-Malmstrom:1966} driven by circularly polarized light; whereas the symmetric part characterizes the ICME induced by linearly polarized light.  Since circularly polarized light breaks time-reversal symmetry, 
a nonvanishing antisymmetric response can exist even in non-magnetic materials. In contrast, the symmetric component  requires the material itself to break time-reversal symmetry, as in ferromagnets. For this reason, this work focus on the ICME arising from the symmetric components.

It is important to note that conventional antiferromagnets are invariant under the combined operation of time reversal with  translation or spatial inversion.  This symmetry forbids linearly polarized light from inducing a static magnetization.  In contrast, altermagnets are invariant under a combined operation of spin flipping and a lattice rotation~\cite{PhysRevX.12.031042}, which can be viewed as an effective time-reversal symmetry. {However, the application of linearly polarized light, which is not invariant under rotational symmetry, breaks the effective time-reversal symmetry of the system}, thereby allowing the ICME to occur in altermagnets.

 To elucidate the distinct features of the ICME in altermagnets, we  consider a $d$-wave altermagnet that has 
$[ C_2||C_{4z}]$ symmetry, where $C_2$ denotes the spin inversion and $C_{4z}$ is four fold spatial rotation around the $z$-axis~\cite{PhysRevX.12.031042}. 
 This symmetry leads to the condition
\begin{align}
	 	 \chi_{ijl}=-R_{jj'}R_{ll'}\chi_{ij'l'},\label{Eq:R}
\end{align}
 where $R$ is the matrix representation of $\pi/2$ rotation about the $z$-axis. Note that the $C_{4z}$ operation does not act on the first index since spin and spatial rotations are decoupled.  
 
 {We consider a geometry in which the light is incident from the $z$-direction and polarized in the $x-y$ plane [see Fig.~\ref{Fig:fig1}(a)]}, 
 then using the constraint Eq.~\eqref{Eq:R},  we find that the diagonal response function satisfies $\chi_{ixx}=-\chi_{iyy}$. The induced magnetization thus takes the  form
\begin{align}
	M_i=(\chi_{ixx}\cos{2\phi}+\chi_{ixy}\sin{2\phi})E^2,
\end{align}
where $\phi$ is the angle between the light's polarization vector and the $x$-axis in the $x-y$ plane, as illustrated in Fig.~\ref{Fig:fig1}(b). 
Notably, the magnetization exhibits an angular dependence with $\pi$ periodicity and changes sign with varying $\phi$. 
This distinct feature is a direct consequence of the unique symmetry of altermagnets and is absent in both ferromagnets and conventional antiferromagnets. {Note that the response function vanishes identically for any rotational symmetry other than the $d$-wave, as enforced by Eq.~\eqref{Eq:R}. A nonzero ICME can emerge in other altermagnetic symmetries under off-normal light incidence, as in this setup the light can couple to an effective $d$-wave-like electronic structure~\cite{JJHe2}.}

\emph{Microscopic theory.\textemdash}
Here we develop a quantitative description of the ICME. 
To reveal the microscopic origin of the effect, we first consider the simplest continuum model for the
$d$-wave altermagnets described by $H=J(k^2_x-k^2_y)\sigma_z$, where $J$ is the altermagnetic coupling constant, $\sigma_z$ is the third Pauli matrix, and $\mathbf{k}$ is the momentum operator. In the presence of light and using the velocity gauge~\cite{PhysRevB.99.045121}, the momentum  is replaced by $\mathbf{k}\to \mathbf{k}+e\mathbf{A}(t)$, where $-e$ is the electric charge and $\mathbf{A}(t)$ is the vector potential corresponding to the electric field, $\mathbf{E}=-\partial_t\mathbf{A}$. Applying  linearly polarized light field $\mathbf{A}=(A_x\mathbf{e}_x+A_y\mathbf{e}_y)\cos(kz-\Omega t)$, we find that the Hamiltonian contains a term $Je^2 (A^2_x-A^2_y)[\cos2(kz-\Omega t)+1]\sigma_z/2$. The time-dependent  part is responsible for the second harmonic generation that is not of our current interest. The time independent part acts as an effective  magnetic field. The Zeeman splitting leads to the familiar Pauli paramagnetism and gives rise to a static magnetization with a simple $1/\Omega^2$ frequency dependence.
{We mention that calculating the Pauli paramagnetic susceptibility does not require explicit spin-flip terms in the Hamiltonian~\cite{AM_solidstate}.} 

\begin{figure}[htb]
	\includegraphics[angle=0,width=0.75\linewidth]{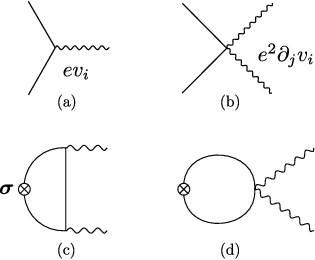}
	\caption{Electron-photon coupling vertexes [(a) and (b)] and Feynman diagrams illustrating the light induced magnetization [(c)-(d)]. The solid lines represent electron's Green's functions, the wavy lines correspond to the vector potential, and the cross denotes Pauli matrices.}
	\label{Fig:feynman}
\end{figure}

The above arguments provide an intuitive explanation of the ICME in altermagnets in terms of the Pauli paramagnetism. Now we develop a general microscopic  theory of the ICME in the independent electron approximation. 
The single particle Hamiltonian is $H(\mathbf{k})$, which becomes  $H(\mathbf{k}+e\mathbf{A})$ in the presence of light~\cite{PhysRevB.99.045121}. As we are interested in the second order effect of the light, it is necessary to expand the perturbed Hamiltonian up to the second order of the vector potential. This  leads to the following perturbed Hamiltonian
\begin{align}\label{Eq:H_perturb}
	H(\mathbf{k}+e\mathbf{A})\approx H(\mathbf{k})+ev_iA_i+\frac{e^2}{2}\partial_j v_i A_iA_j,
\end{align} 
where $v_i=\partial_{i} H(\mathbf{k})$ with $\partial_i=\partial_{k_i}$ is the velocity operator and $\partial_j v_i$ gives rise to the diamagnetic response. 
 The corresponding interacting vertexes are shown in Figs.~\ref{Fig:feynman}~(a)-(b).

We employ the Keldysh formalism~\cite{HAU2008} to calculate the induced magnetization, which can be expressed as
\begin{align}\label{Eq:M}
	\mathbf{M}=
	i\mu_B\tr \bm{\sigma}\delta G^{<},
\end{align}
where $\mathrm{tr}$ denotes the trace in the spin space,  $\mu_B$ 
is the Bohr magneton, $\bm{\sigma}$ are the Pauli matrices, and $\delta G^<$ is the second order correction to the nonequilibrium lesser Green's function induced by  the applied perturbation. The lesser Green's function can be obtained from the time ordered Green's function on the Keldysh contour through the Langreth theorem~\cite{HAU2008}.  The physical processes contributing to the magnetization are represented diagrammatically in Figs.~\ref{Fig:feynman}~(c)-(d). 
The expression Eq.~\eqref{Eq:M} contains both static magnetization and the second harmonic generation. Extracting the static component, we derive  the response function characterizing the ICME
\begin{widetext}
	\begin{align}\label{Eq:Chi_ijl}
		\chi_{ijl}
		=
		&\frac{e^2\mu_B}{4\Omega^2}\Im\sum_{q_0=\pm\Omega}\int\frac{\mathrm{d}\omega}{2\pi}\frac{\mathrm{d}\mathbf{k}}{(2\pi)^d}\trace{\sigma_i} G^r(\omega,\mathbf{k})v_j\qty(\mathbf{k})[G^r(\omega+q_0,\mathbf{k})-G^r(\omega,\mathbf{k})]v_l\qty(\mathbf{k})G^r(\omega,\mathbf{k})f(\omega)\nonumber\\
		&+\frac{e^2\mu_B}{4\Omega^2}\Im \sum_{q_0=\pm\Omega}\int\frac{\mathrm{d}\omega}{2\pi}\frac{\mathrm{d}\mathbf{k}}{(2\pi)^d}\trace{\sigma_i} G^r(\omega,\mathbf{k})v_j\qty(\mathbf{k})G^r(\omega+q_0,\mathbf{k})v_l\qty(\mathbf{k})G^a(\omega,\mathbf{k})[f(\omega+q_0)-f(\omega)]+(j\leftrightarrow l),
	\end{align}
\end{widetext}
here $d$ is the spatial dimension,  $G^{r}$ and $G^{a}$ are the unperturbed retarded and advanced Green's functions, respectively, and  $f(\omega)=1/[e^{\omega/(k_B T)}+1]$ is the Fermi-Dirac distribution with $T$ being the temperature and $k_B$ being the Boltzmann's constant. 
The second term in the square bracket in the first line of Eq.~\eqref{Eq:Chi_ijl} corresponds to Fig.~\ref{Fig:feynman}~(d) and the other terms come from Fig.~\ref{Fig:feynman}~(c). 
As can be seen from the expression, the diamagnetic contribution is essential for obtaining a finite result in the zero-frequency limit.

Our result Eq.~\eqref{Eq:Chi_ijl} is independent of the specific form of the  Hamiltonian. For the independent electron approximation, the Green's function can be written as
\begin{align}
	G^{r/a}(\omega,\mathbf{k})=\sum_{s}\frac{|u_s(\mathbf{k})\rangle \langle u_s(\mathbf{k})|}{\omega -\xi_{s,\mathbf{k}}\pm i\Gamma},
\end{align}
where $|u_s(\mathbf{k})\rangle$ is the $s$-th eigenstate of the single particle Hamiltonian with the eigenenergy $\epsilon_{s,\mathbf{k}}$, i.e.,  $H(\mathbf{k})|u_s(\mathbf{k})\rangle=\epsilon_{s,\mathbf{k}}|u_s(\mathbf{k})\rangle$, and $\xi_{s,\mathbf{k}}=\epsilon_{s,\mathbf{k}}-\mu$ with $\mu$ being the chemical potential. We have introduced a constant decay rate $\Gamma$ to phenomenologically describe the effect of spectral broadening due to impurities~\cite{nonlinearoptics} and the clean limit is recovered by taking $\Gamma\to 0$.

\begin{figure}
	\includegraphics[angle=0,width=0.8\linewidth]{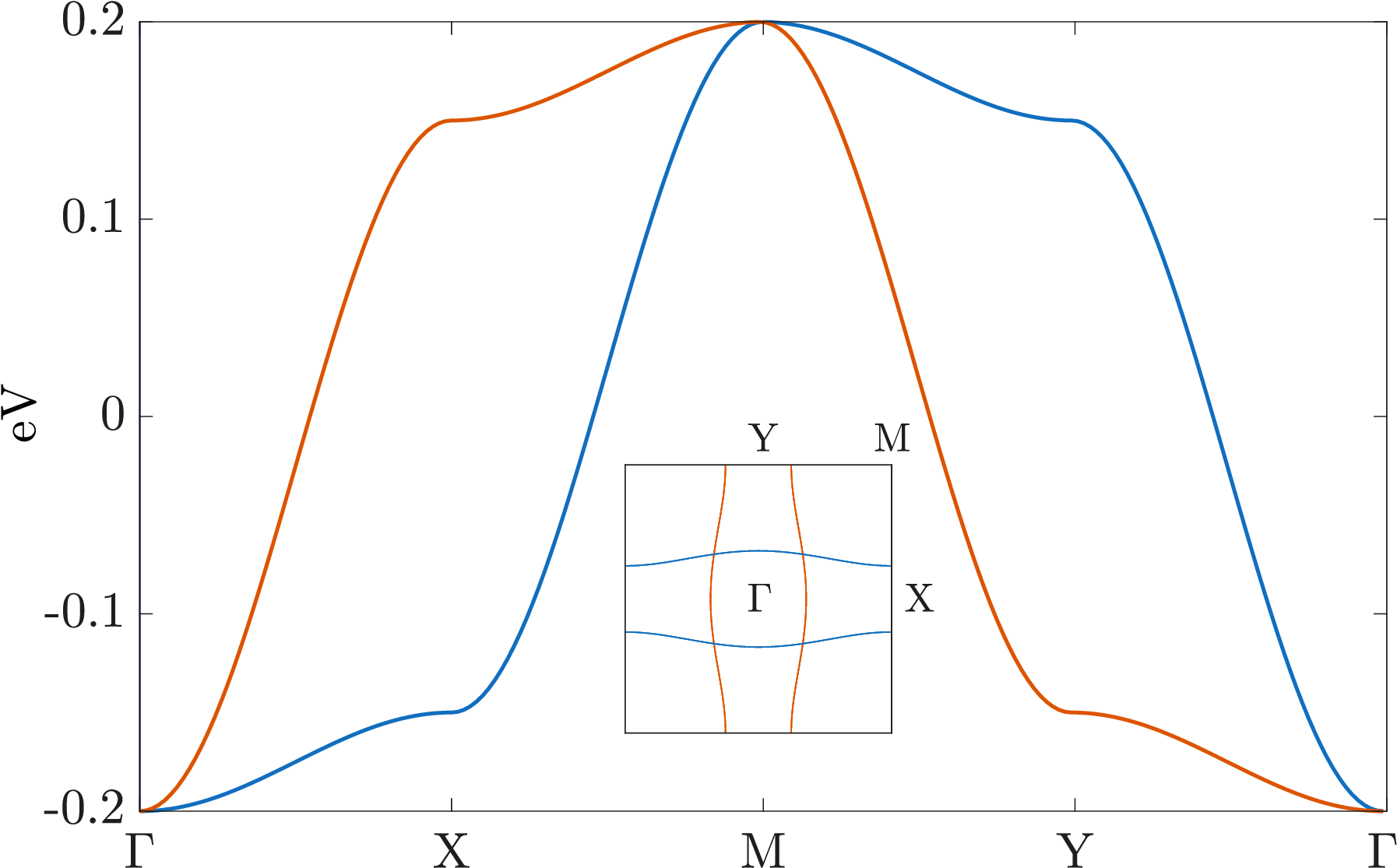}
	\caption{The energy dispersion along high symmetry lines corresponding the model Eq.~\eqref{Eq:HamAlterM}. 
		Different colors represent opposite spin states.
		The inset shows the Fermi surfaces. 
		The hopping parameters are $t=$\SI{0.1}{eV} and $t_J=$\SI{0.075}{eV}, and the chemical potential is taken to be $\mu=$\SI{-0.1}{eV}, corresponding to the parameters for KRu$_4$O$_8$~\cite{PhysRevX.12.031042}. }
	\label{Fig:dispersion}
\end{figure}

\emph{Applications.\textemdash}
As a representative  example,  we consider a planar $d$-wave altermagnet described by the Hamiltonian~\cite{PhysRevX.12.031042}
\begin{align}
	H(\mathbf{k})=h_0(\mathbf{k})+h_{\mathrm{a}}(\mathbf{k})\hat{\mathbf{n}}\cdot\bm{\sigma},\label{Eq:HamAlterM}
\end{align}
where $h_0=-t(\cos{k_x}+\cos{k_y})$ is the usual kinetic energy,  $h_{\mathrm{a}}=-t_J(\cos{k_x}-\cos{k_y})$ describes a $d$-wave altermagnet, and $\hat{\mathbf{n}}$ is a unit vector characterizing the direction of the N\'eel order. The lattice constant is taken to be unity. 
The order parameter exhibits $d_{x^2-y^2}$-wave symmetry, which is related to $d_{xy}$-wave symmetry by a $\pi/4$ rotation.  
The energy dispersion is $\epsilon_\eta=h_0+\eta h_a$ with $\eta=\pm$ denoting the two bands with opposite spins. 
This Hamiltonian represents a minimal model for $d$-wave altermagnetic metal~\cite{PhysRevX.12.031042,weber2024ultrafast} KRu$_4$O$_8$. The parameters are~\cite{PhysRevX.12.031042} $t=$\SI{0.1}{eV}, $t_J=$\SI{0.075}{eV}, and $\mu=$\SI{-0.1}{eV}. Figure~\ref{Fig:dispersion} displays the energy dispersions along high symmetry lines, with opposite spin states indicated by different colors. The corresponding Fermi surfaces are shown in the inset. {The spin-orbit coupling in  KRu$_4$O$_8$ is negligible~\cite{weber2024ultrafast}, and spin point group analysis~\cite{Etxebarria_2025,elcoro2026} of the material permits only $\chi_{ixy}$ and the relation $\chi_{ixx} = -\chi_{iyy}$ with $i$ being the direction of the N\'eel order, while all other components are forbidden. The effect of spin-orbit coupling is further discussed in  Supplemental Material~\cite{SI}.}
\begin{figure*}
	\includegraphics[angle=0,width=0.9\linewidth]{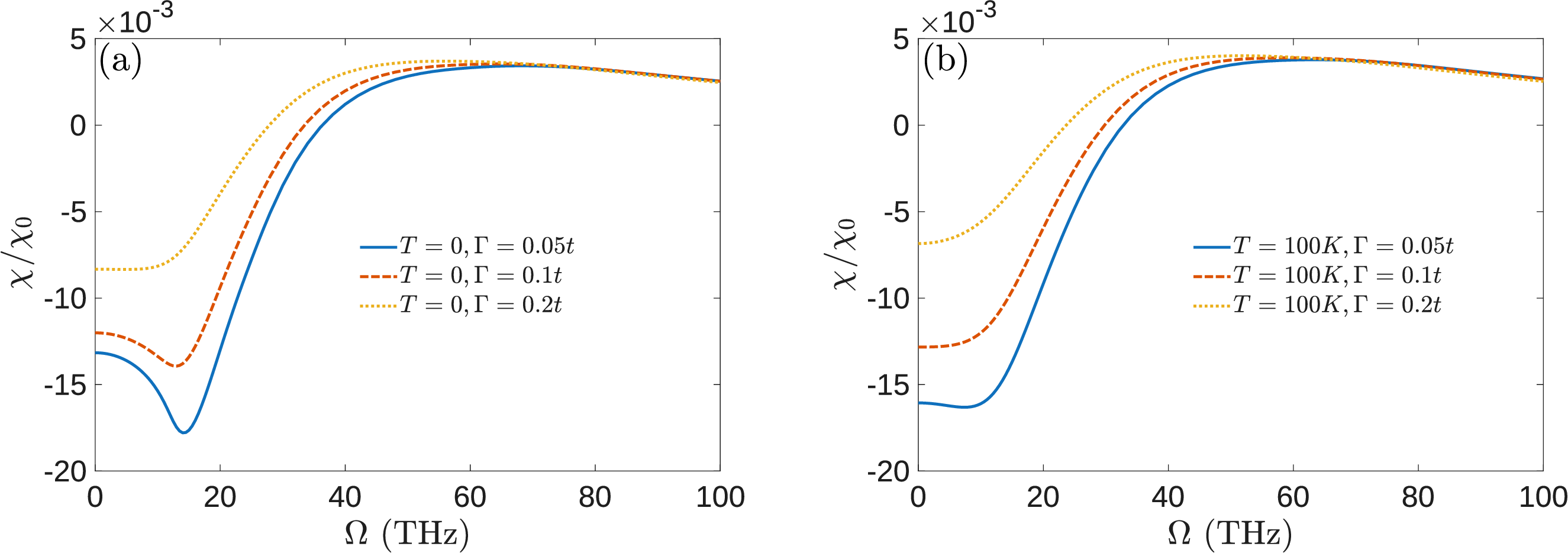}
	\caption{The dimensionless response $\chi/\chi_0$ as a function of light frequency $\Omega$ for different decay rates at zero temperature (a) and $T=$\SI{100}{K} (b), where $\chi_0=e^2\mu_B/t^2$. Other parameters are the same as those in Fig.~\ref{Fig:dispersion}. 
	}
	\label{Fig:results}
\end{figure*}

Substituting the corresponding Green's function into the expression Eq.~\eqref{Eq:Chi_ijl},  the response function can be simplified as
\begin{align}
	\chi_{ijl}=&e^2\mu_B \hat{n}_i\Im\sum_{\eta=\pm}\nonumber\\
	&\int 
	\frac{\mathrm{d}\mathbf{k}\mathrm{d}\omega}{(2\pi)^3}
	\frac{\eta \partial_j\xi_\eta\partial_l\xi_\eta f(\omega)}{(\omega-\xi_\eta +i\Gamma)^3[(\omega-\xi_\eta+i\Gamma)^2-\Omega^2]}.
\end{align}
Note that the induced magnetization is along the direction of the N\'eel order, consistent with the spin group analysis. This provides an optical way to detect the N\'eel order  in altermagnets. We mention that the response function in the static limit remains nonzero, giving rise to a dc nonlinear magnetoelectric effect~\cite{trama2024,PhysRevB.110.184407}.

Now we present the numerical results of the response function.  
The model Hamiltonian Eq.~\eqref{Eq:HamAlterM} possesses a mirror symmetry $M_x$, and consequently, the component $\chi_{ixy}$ vanishes identically. The only independent component is $\chi_{ixx}\equiv \hat{n}_i\chi$. 
Figure~\ref{Fig:results} shows the dimensionless response function $\chi/\chi_0$, with $\chi_0=e^2\mu_B/t^2$, as a function of light frequency for several different decay rates at $T=0$ [Fig.~\ref{Fig:results}(a)] and $T=$\SI{100}{K} [Fig.~\ref{Fig:results}(b)]. 
As shown in the figure, for small decay rates $\Gamma$, the response function shows a minimum with increasing light frequency. The depth of this minimum decreases with rising temperature and decay rate. 
For sufficiently large $\Gamma$ or high temperature,  the minimum disappears. Further increasing frequency, the response function changes sign. In the high frequency limit, the response function  becomes largely insensitive to both temperature and decay rate. To check the robustness of the effect against impurity scattering, we also calculate the response function using an energy-dependent decay rate within the Born approximation and obtain similar results, see  Supplemental Material~\cite{SI}.

To estimate the order of magnitude of the induced magnetization, we consider a laser beam with an electric strength of  $E\sim $\SI{1}{MeV/cm}~\cite{Huisman_2016},  corresponding to a light intensity of \SI{1.3}{GW/cm^2}. This yields a scale of $\chi_0 E^2  \sim \mu_B$\si{nm^2}. Given that the lattice constant of KRu$_4$O$_8$ is approximately $10$\AA~\cite{weber2024ultrafast}, the induced spin angular momentum per unit cell is estimated as  $\chi/\chi_0\mu_B$ per unit cell. For $\Omega=$\SI{10}{THz} and $\Gamma=$\SI{0.01}{eV}, this results in a value of approximately $-0.016\mu_B$ per unit cell at \SI{100}{K}.

Since $\chi_{ixy}$ vanishes, 
the induced magnetization is largest when the polarization angle is $\phi=0$ or $\pi$. 
If the altermagnetic order has $d_{xy}$-wave symmetry instead of $d_{x^2-y^2}$-wave,  then $\chi_{ixx}$ vanishes while $\chi_{ixy}$ becomes nonzero, shifting the maxima of the magnetization to $\phi=\pi/4$ and $\phi=3\pi/4$. In the cases where these two order parameters coexist, the maximal magnetization occurs at intermediate angles. Consequently, 
the angular dependence of the magnetization serves as a probe of the underlying altermagnetic symmetry. {The predicted effect can be detected using the pump-probe setup~\cite{weber2024optical}, and the induced magnetization can be measured using established techniques such as MOKE or SQUID. Our numerical estimate indicates that the effect is stronger in clean samples with low scattering rate and at low temperatures.} We note that, in a recent experiment on a RuO$_2$ thin film~\cite{weber2024optical},
optical excitation of the magnetization has been studied using the pump-probe technique, and a maximal Kerr rotation was observed at $\phi=\pi/4$ and $3\pi/4$, indicating an altermagnetic order with $d_{xy}$-wave symmetry.

\emph{Conclusion.\textemdash}In this work, we present a theoretical study of the ICME in $d$-wave altermagnets. 
Employing the Keldysh formalism, we develop a microscopic theory of the phenomenon. Our approach, which combines symmetry analysis with this microscopic framework, reveals that the N\'eel vector governs the direction of the induced magnetization, while its magnitude exhibits a sinusoidal dependence on the light's polarization angle—a hallmark signature of the system's unique symmetry. 
{Although we employ the single-particle approximation in our microscopic theory, the robustness of our predication for  materials belonging to the $d$-wave altermagnet class is ensured by symmetry. Interaction effects can be incorporated into the single-particle Hamiltonian through first principle calculations or mean-field approximation~\cite{Leeb_2024}.}
As a concrete application, we evaluate the effect for the candidate altermagnet KRu$_4$O$_8$, estimating an induced magnetization on the order of $1\%\mu_B$ per unit cell. 
{Our findings highlight the ICME as a tool for the optical {detection} of  magnetization in {$d$-wave} altermagnets. This provides a method for the imaging of magnetic domains~\cite{WKuch}. However, as the induced magnetization is collinear with the N\'eel vector, it cannot generate the torque for switching. For that purpose, circularly polarized light whose angular momentum can induce a non-collinear moment via the inverse Faraday effect could be employed.}

	{\it Acknowledgments.\textemdash} 
	We are grateful to 
	Baotao Fu and Deping Guo for useful discussions. This work was supported by National Natural Science
	Foundation of China under Grants No.~12204329 and No.~12204331.

\bibliography{ICME.bib}

\setcounter{equation}{0}
\setcounter{figure}{0}
\setcounter{table}{0}

\renewcommand{\theequation}{S\arabic{equation}}
\renewcommand{\thefigure}{S\arabic{figure}}
\renewcommand{\thetable}{S\arabic{table}}
\renewcommand{\bibnumfmt}[1]{[S#1]}
\newcommand{\beginsupplement}{%
	\setcounter{equation}{0}
	\renewcommand{\theequation}{S\arabic{equation}}%
	\setcounter{table}{0}
	\renewcommand{\thetable}{S\arabic{table}}%
	\setcounter{figure}{0}
	\renewcommand{\thefigure}{S\arabic{figure}}%
	\setcounter{section}{0}
	\renewcommand{\thesection}{S\Roman{section}}%
	\setcounter{subsection}{0}
	\renewcommand{\thesubsection}{S\Roman{section}.\Alph{subsection}}%
}

\clearpage
\pagebreak
\widetext

\begin{center}
	\textbf{\large Supplemental material to ``Nonlinear opto-magnetic signature of $d$-wave altermagnets"}
\end{center}

	\section{The effect  of spin-orbit coupling}

\begin{figure}[h]
	\includegraphics[width=0.8\textwidth]{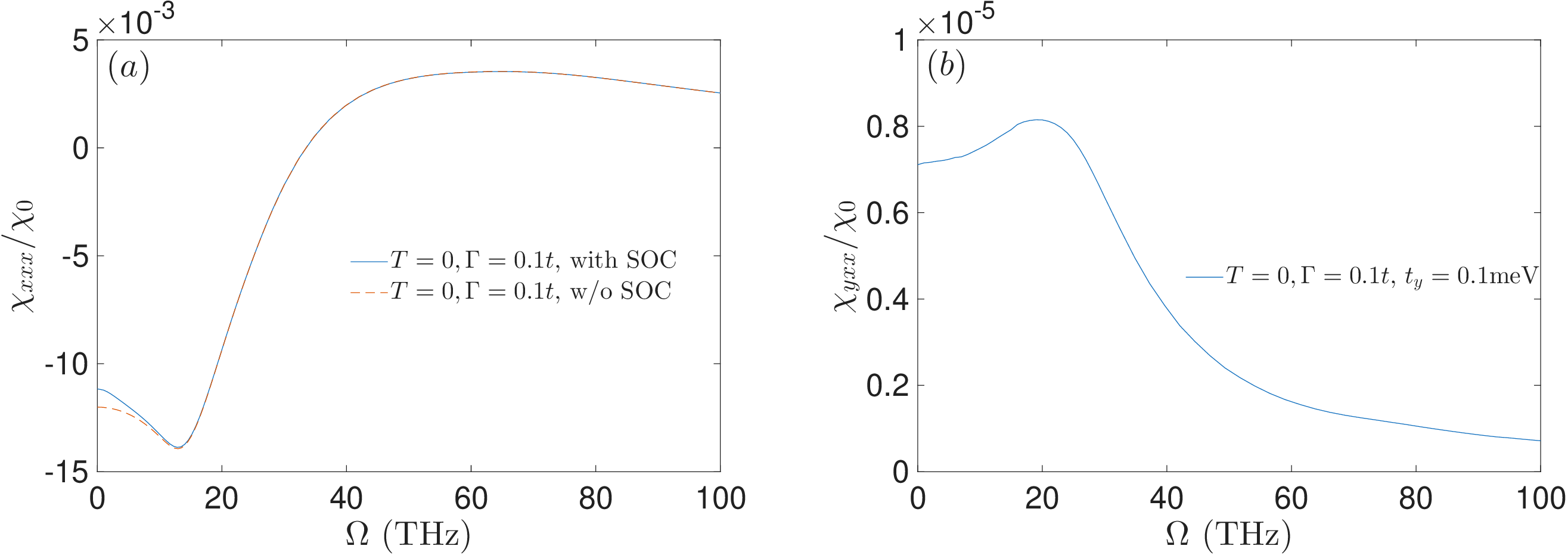}
	\caption{Effects of spin-orbit couplings on the response functions. (a): The response function $\chi_{xxx}$  in the presence of inter-sublattice spin-orbit coupling. The Hamiltonian is Eq.~\eqref{Eq:HamSOC}. We find that $\chi_{xyy}=-\chi_{xxx}$, and other components vanish. Therefore the magnetization is along the $x$-direction, which is the direction of N\'eel order. The spin-orbit coupling modifies $\chi_{xxx}$ slightly. (b): The same-sublattice spin-orbit coupling [Eq.~\eqref{Eq:Ham_hy}] induces nonzero $\chi_{yxx}=\chi_{yyy}$, which is much smaller than $\chi_{xxx}$.}\label{Fig:withSOC1}
\end{figure}

In this section we study the effect of the spin-orbit coupling. According to Refs.~\cite{PhysRevX.12.031042,weber2024ultrafast}, the magnetic order of KRu$_4$O$_8$ is along the $x$-direction, and the magnetic point group is $2'/m'$. There are thus 10 independent components~\cite{elcoro2026,Birss}: 
$\chi_{xxx}$, $\chi_{xyy}$, $\chi_{xxy}$, $\chi_{yxx}$, $\chi_{yyy}$, $\chi_{yxy}$, $\chi_{xzz}$, $\chi_{yzz}$, $\chi_{zxz}$, and $\chi_{zyz}$. To calculate these components, we use the four-band model~\cite{weber2024ultrafast},
\begin{align}\label{Eq:HamSOC}
	H_0=(h_0+J)\tau_0\otimes\sigma_0+h_a\tau_z\otimes\sigma_0+J\tau_z\otimes\sigma_x+h_{z}\tau_y\otimes\sigma_z,
\end{align}
where $\bm{\tau}$ are the Pauli matrices in the sublattice space and $\bm{\sigma}$ are the Pauli matrices in the spin space, $\tau_0$ and $\sigma_0$ are identity matrices,   $h_0=-t(\cos{k_x}+\cos{k_y})$,  
$h_{\mathrm{a}}=-t_J(\cos{k_x}-\cos{k_y})$,
$h_{z}=-4t_z\sin{\frac{k_x}{2}}\sin{\frac{k_y}{2}}$, and $J$ is the exchange field  in the $x$-direction. The spin-orbit coupling is described by the $h_z$ term, which is an inter-sublattice spin flipping term.
The parameters are~\cite{PhysRevX.12.031042, weber2024ultrafast} $t=$\SI{0.1}{eV}, $t_J=$\SI{0.075}{eV}, $J=$\SI{0.2}{eV}, and $t_z=$\SI{2}{meV}. Note that the spin-orbit coupling strength is much smaller than other energy scales. We find that the only non-zero components are $\chi_{xxx}$ and $\chi_{xyy}$, and within our numerical accuracy, $\chi_{xxx} = -\chi_{xyy}$. The result for $\chi_{xxx}$ is shown in  Fig.~\ref{Fig:withSOC1}(a), where we also plot the corresponding result in the absence of spin-orbit coupling. As shown in the figure, the spin-orbit coupling only slightly modifies the value of $\chi_{xxx}$ in the low-frequency limit.

Note that the induced  magnetization remains along the direction of N\'eel order for the model described by Eq.~\eqref{Eq:HamSOC}. This is because the spin-orbit coupling is an inter-sublattice flip term, and in the basis
\begin{align}\label{Eq:Basis}
	\{|\phi_{A}\rangle|\leftarrow\rangle, |\phi_B\rangle|\rightarrow\rangle, |\phi_B\rangle|\leftarrow\rangle,|\phi_A\rangle|\rightarrow\rangle \},
\end{align}
where $|\leftarrow\rangle$ and $|\rightarrow\rangle$ are the eigenstates of $\sigma_x$, the Hamiltonian becomes block diagonalized~\cite{weber2024ultrafast},
\begin{align}
	H_0=\left[\begin{array}{cccc}
		h_0+2J+h_a	& t_z & 0 & 0 \\
		t_z	& h_0+2J-h_a & 0 & 0 \\
		0	& 0 & h_0+h_a & -t_z \\
		0	& 0 & -t_z & h_0-h_a
	\end{array}\right],
\end{align}  
while the spin operator $\sigma_y$ is off-block-diagonal in this basis, so its expectation value vanishes even in the presence of light.

We mention that the tight-binding models for altermagnetic material candidates in Refs.~\cite{Attias_2024,Roig_2024} share a similar structure as the Hamiltonian Eq.~\eqref{Eq:HamSOC}: the spin-orbit coupling couples nearest-neighbor  spins. Therefore, the induced magnetization calculated using the models in Refs.~\cite{Attias_2024,Roig_2024} remains along the direction of the N\'eel order even when spin-orbit coupling is taken into account. 
To go beyond this, we can add to the model Eq.~\eqref{Eq:HamSOC} an intrasublattice spin-orbit coupling  term,
\begin{align}\label{Eq:Ham_hy}
	h_{y}=-t_y(\cos{k_x}+\cos{k_y})\tau_0\sigma_y.
\end{align}
This term opens a gap at the $\Gamma$ point. Since DFT calculations do not observe such a  gap~\cite{weber2024ultrafast}, $t_y$ must be very small, thus we take $t_y=$\SI{0.1}{meV}. We find that $\chi_{xxx}$ is largely unaffected, while $\chi_{yxx}$ becomes nonzero with $\chi_{yyy} = \chi_{yxx}$. The numerical result for $\chi_{yxx}$ is shown in  Fig.~\ref{Fig:withSOC1}(b). As one can see, $\chi_{yxx}$ is much smaller than $\chi_{xxx}$, so the magnetization is dominated by $\chi_{xxx}$.

\section{Born approximation for the impurity scattering}

\begin{figure}[h]
	\includegraphics[angle=0,width=0.85\linewidth]{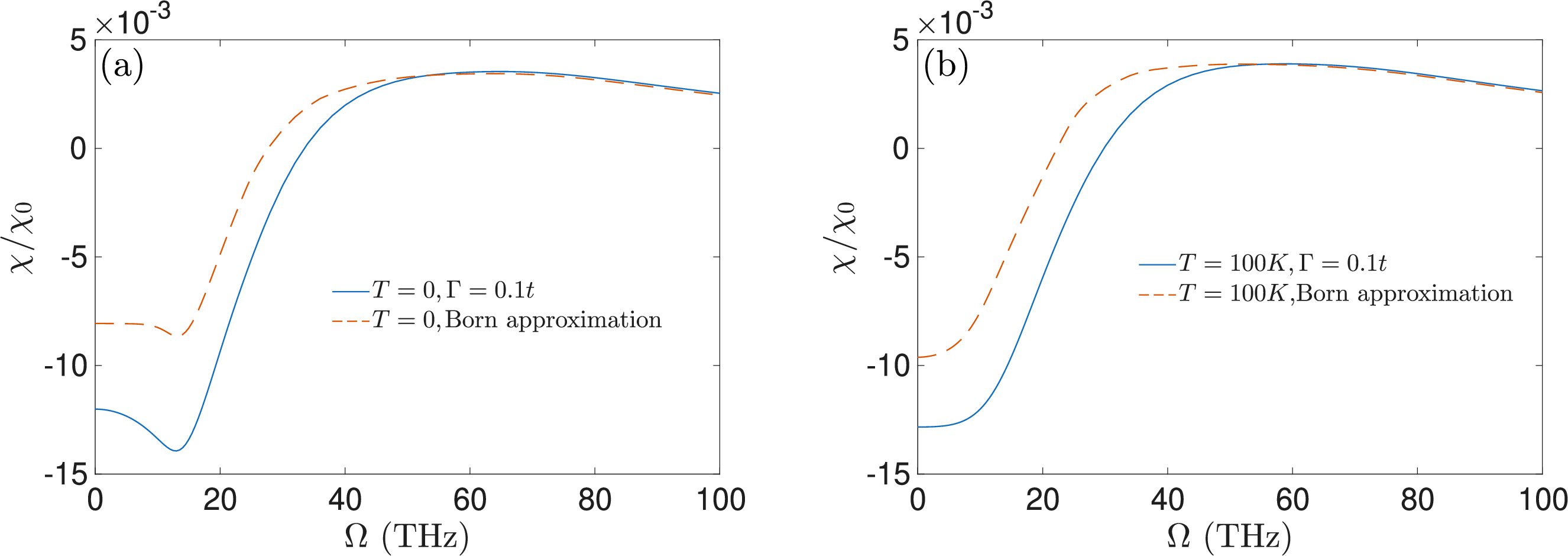}
	\caption{Response function versus frequency at (a) $T = 0$\si{K} and (b) $T = 100$\si{K}, comparing the Born approximation (red dashed line) with the constant decay approximation (blue solid line). In the Born approximation, the parameters are chosen such that the decay rate at the Fermi energy is $\Gamma(\mu)=0.1t$.
	}
	\label{Fig:BornApprox}
\end{figure}
In the main paper we present the results for the response function calculated using  several constant scattering rates. To check the robustness of effect, here we calculate the response function using the scattering rate within the Born approximation. Assuming that the impurity-induced interaction can be modeled by a spin independent $\delta$-function potential, 
\begin{align}
	g\sum_{i}\delta(\mathbf{r}-\mathbf{R}_i),
\end{align}		
where $\mathbf{R}_i$ are random impurity positions and $g$ characterize the interaction strength. 
The scattering rate within the Born approximation is proportional to the impurity density $n_{\mathrm{imp}}$ and the electronic density of states. 
As shown in Fig.~\ref{Fig:BornApprox}, the results obtained using the energy-dependent scattering rate  agree with those from the constant-decay approximation, which underscores the robustness of our findings. 

\end{document}